\documentclass[12pt]{iopart}
\usepackage{stmaryrd}
\usepackage{epsfig}
\usepackage{epstopdf}
\usepackage{txfonts}
\usepackage{amssymb}
\usepackage{cite}
\usepackage{subfigure}
\usepackage{txfonts}
\usepackage{amssymb}
\usepackage{tipa}
\usepackage{graphicx}
\usepackage{txfonts}
\usepackage{amssymb}

\begin{document}

\title[An efficiently and tunable switch between  slow- and fast- light in double mode with an
optomechanical system]{An efficiently  and tunable switch between
slow- and fast- light in double mode with an optomechanical system}

\author{Peng-Cheng Ma,$^{1,*}$\footnote[0]{$^{*,\dag}$Author to whom any correspondence should be addressed.}, Lei-Lei Yan,$^{2}$  Gui-Bin Chen,$^{1}$ Xiao-Wei Li$^{1}$   and You-Bang Zhan$^{1,\dag}$}

\address{$^1$ School of Physics and Electronic Electrical Engineering, Huaiyin Normal University, Huaian 223300, China \\
$^2$ State Key Laboratory of Magnetic Resonance and Atomic and
Molecular Physics, Wuhan Institute of Physics and Mathematics,
Chinese Academy of Sciences, Wuhan 430071, China}
\ead{mapch95812@163.com; ybzhan@hytc.edu.cn}

\begin{abstract}
We study the dynamics of a driven optomechanical cavity coupled to a
charged nanomechanical resonator via Coulomb interaction. We find
the tunable   switch between slow- and fast- light in double-mode
can be observed from the output field by  adjusting the laser-cavity
detuning  in this coupled system. Moreover, the requencies of signal
light can be tuned by Coulomb coupling strength. The proposal may
have potential application in optical communcation and nonlinear
optics.
\end{abstract}

\maketitle

\section{Introduction}

The control of slow and fast light propagation  is a challenging
task. Research on slow- and fast- light systems has increased from
both theoretical and experimental aspects in physics
\cite{science326-1074,prl74-2447}. The first superluminal light
propagation was observed in a resonant system \cite{prl48-738},
where the laser propagates without appreciable shape distortion but
experiences very strong resonant absorption. Various techniques have
been developed to realized slow and fast light in atomic vapors
\cite{nature397-594,prl82-5229,prl83-1767} and solid materials
\cite{prl88-023602,prl90-113903}. To reduce absorption, most of
those works \cite{nature397-594,prl82-5229,prl83-1767,prl88-023602}
are based on the electromagnetically induced transparency (EIT) or
coherent population oscillation (CPO) \cite{ol29-2291}.

    On the other hand,  opto-mechanical \cite{nature444-67,natue444-71,natue444-75,prl97-243905,nature460-724,nature478-89}
systems   have advanced rapidly, which are promising candidates for
realizing architectures exhibiting quantum behavior in macroscopic
structures.  Also, a lot of them have been demonstrated
experimentally in this systems, for example, quantum information
transfer \cite{science338-1609}, normal mode splitting
\cite{nature460-724,prl101-263602}, optomechanically induced
transparency (OMIT) \cite{science330-1520}, frequency transfer
\cite{natcommun3-1196}. Slow- and fast- light also has been
successfully observed in this system \cite{nature472-69}, where an
optically tunable delay of 50 ns with near-unity optical
transparency and superluminal light  with a 1.4-$\mu$s signal
advance. Most recently proposals are proposed in this system, such
as slow light based on an optomechanical cavity with a Bose-Einstein
condensate (BEC) \cite{pra83-055803} and fast light in reflection
meanwhile slow light in transmission \cite{pra87-013824}. Moreover,
we have demonstrated an efficient switch between slow and fast light
in microwave regime \cite{oe22-3621}.

    However, above proposals and experiments all are work in one
optical mode. In this paper, we theoretically investigate the slow-
and fast- light in double light mode based on the optomechanical
system. Compare to resent proposal
\cite{nature472-69,pra83-055803,pra87-013824}, we can efficiently
switch from slow- to fast- light in double light mode only by
adjusting the effective laser-cavity detuning in reflective light.
Moreover, the frequency of  output light can be tunable according to
Coulomb coupling strength

This paper is structured as follows. In Sec. 2 we present the model
and the analytical expressions of the optomechanical system and
obtain the solutions. Sec. 3 includes numerical calculations for the
efficiently  double mode and tunable switch from slow to fast based
on recent experimental parameters.  The last section is a brief
conclusion.

\section{Model and  solutions}

\begin{figure}[tbp]
\centering
\includegraphics[width=8cm,height=5cm]{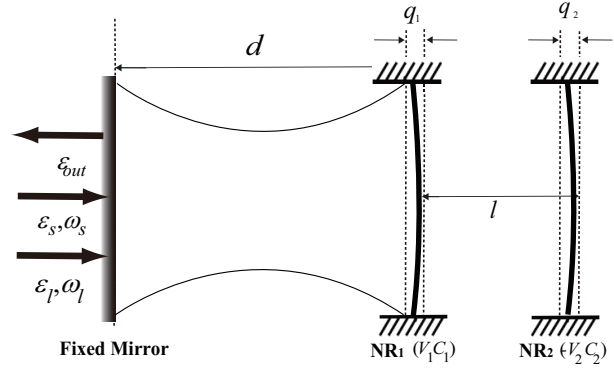}
\caption{(Color online) Schematic diagram of the system. A
high-quality Fabry-P\'{e}rot cavity consists of a fixed mirror and a
movable mirror NR$_{1}$. NR$_{1}$ is charged by the bias gate
voltage $V_{1}$ and subject to the Coulomb force due to another
charged NR$_{2}$ with the bias gate voltage $-V_{2}$. The
optomechanical cavity of the length $d$ is driven by two light
fields, one of which is the pump field $\protect\varepsilon_{l}$
with frequency $\protect\omega_{l}$ and the other of which is the
probe field $\protect\varepsilon_{s}$ with frequency
$\protect\omega_{s}$. The output field is represented by
$\protect\varepsilon_{out}$. $q_{1}$ and $q_{2}$ represent the small
displacements of NR$_{1}$ and NR$_{2}$ from their equilibrium
positions, with $l$ is the equilibrium distance between the two
NRs.}
\end{figure}
We begin with the Hamiltonian of the opto-mechanical system. As
shown schematically in  Fig. 1, the Hamiltonian is given by
\cite{ma},
\begin{eqnarray}
&&H=\hbar \omega _{a}a^{\dag
}a+(\frac{p_{1}^{2}}{2m_{1}}+\frac{1}{2}
m_{1}\omega _{1}^{2}q_{1}^{2})  \nonumber \\
&&+(\frac{p_{2}^{2}}{2m_{2}}+\frac{1}{2}m_{2}\omega
_{2}^{2}q_{2}^{2})-\hbar
ga^{\dag }aq_{1}+\hbar\lambda q_1q_2  \nonumber \\
&&+i\hbar \varepsilon _{l}(a^{\dag }e^{-i\omega _{l}t}-H.c.)+i\hbar
(a^{\dag }\varepsilon _{s}e^{-i\omega _{s}t}-H.c.),
\end{eqnarray}
where the first term is for the single-mode cavity field with
frequency $\omega_{a}$ and annihilation (creation) operator $a\
(a^{\dag })$. The second (third) term describes the vibration of the
charged NR$_{1}$ (NR$_{2}$ ) with frequency $\omega_{1}$
($\omega_{2}$), effective mass $m_{1}$ ($ m_{2}$), position $q_{1}$
($q_{2}$) and momentum operator $p_{1}$ ($p_{2}$)
\cite{pra85-021801}. The forth term denote the NR$_{1}$ couples to
the cavity field due to the radiation pressure with the coupling
strength $g=\frac{\omega_{a}}{d}$ with $d$ being the cavity length.

The fifth term  presents the Coulomb coupling between the charged
NR$_{1}$ and  the charged NR$_{2}$
 and $\lambda
=\frac{C_{1}V_{1}C_{2}V_{2}}{2\pi \hbar \varepsilon _{0}r_{0}^{3}}$
\cite{ma,pra72-041405,prl93-266403}, where the NR$_{1}$ and NR$_{2}$
take the charges $C_{1}V_{1}$ and $-C_{2}V_{2}$, with $C_{1}(C_{2})$
and $V_{1}(-V_{2})$ being the capacitance and the voltage of the
bias gate, respectively. $l$ is the equilibrium distance between the
two NRs. $q_{1}$ and $q_{2}$ represent the small displacements of
NR$_{1}$ and NR$_{2}$ from their equilibrium positions,
respectively. The last two terms in Eq. (1) describe the
interactions between the cavity field and the two input fields,
respectively. The strong (week) laser (signal) field owns the
frequency $\omega_{l}$ ($\omega_{s}$) and the amplitude $
\varepsilon_{l}=\sqrt{2\kappa\wp_{l}/\hbar\omega_{l}}$ ($\varepsilon
_{s}= \sqrt{2\kappa \wp _{s}/\hbar\omega _{s}}$), whit $\wp_{l}$
($\wp _{s}$) is the power of the laser (signal) field and $\kappa$
is the cavity decay rate.

In a frame rotating with the frequency $\omega_l$ of the laser
field, the Hamiltonian of the system Eq.(1) can be rewritten as,
\begin{eqnarray}
&&H=\hbar\Delta_aa^\dag a
+(\frac{p_1^2}{2m_1}+\frac{1}{2}m_1\omega_1^2q_1^2)
\nonumber \\
&&+(\frac{p_2^2}{2m_2}+\frac{1}{2}m_2\omega_2^2q_2^2) -\hbar ga^\dag
aq_1
+\hbar \lambda q_1q_2  \nonumber \\
&&+i\hbar\varepsilon_l(a^\dag -a)
+i\hbar(a^\dag\varepsilon_se^{-i\delta t}-H.c.),
\end{eqnarray}
where $\Delta_a=\omega_a-\omega_l$ is the detuning of the laser
field from the bare cavity, and $\delta=\omega_s-\omega_l$ is the
detuning of the singal field from the laser field.

Considering photon losses from the cavity , we may describe the
dynamics of the system governed by Eq. (2) using following nonlinear
quantum Langevin equations \cite{pra85-021801},
\begin{eqnarray}
&&\langle \dot{q_{1}}\rangle =\frac{\langle p_{1}\rangle }{m_{1}},
\nonumber
\label{7} \\
&&\langle \dot{p_{1}}\rangle =-m_{1}\omega _{1}^{2}\langle
q_{1}\rangle -\hbar \lambda \langle q_{2}\rangle +\hbar g\langle
c^{\dag }\rangle \langle
c\rangle -\gamma _{1}\langle p_{1}\rangle ,  \nonumber \\
&&\langle \dot{q_{2}}\rangle =\frac{\langle p_{2}\rangle }{m_{2}},  \nonumber \\
&&\langle \dot{p_{2}}\rangle =-m_{2}\omega _{2}^{2}\langle
q_{2}\rangle -\hbar \lambda \langle q_{1}\rangle -\gamma _{2}\langle
p_{2}\rangle ,
\nonumber \\
&&\langle \dot{a}\rangle =-[\kappa +i(\Delta _{a}-g\langle
q_{1}\rangle )]\langle a\rangle +\varepsilon _{l}+\varepsilon
_{s}e^{-i\delta t},
\end{eqnarray}
here $\gamma _{1}$ and $\gamma _{2}$  are the decay rates for NR$
_{1}$ and NR$_{2}$, respectively. Where we have been considered: (i)
The quantum Brownian noise $\xi _{1}$ $(\xi _{2})$ comes from the
coupling between NR$_{1}$ (NR$_{2}$) and its own environment with
zero mean value \cite{pra77-033804}; (ii) $c_{in}$ is the input
vacuum noise operator with zero mean value \cite{pra77-033804} and
under the mean field approximation $\langle Qa\rangle =\langle
Q\rangle \langle a\rangle $ \cite{pra81-041803}. which is a set of
nonlinear equations and the steady-state response in the frequency
domain is composed of many frequency components. We suppose the
solution with the following form \cite{pra86-053806}
\begin{eqnarray}
&&\langle q_{1}\rangle =q_{1s}+q_{1+}\varepsilon _{s}e^{-i\delta
t}+q_{1-}\varepsilon _{s}^{\ast }e^{i\delta t},  \nonumber    \\
&&\langle p_{1}\rangle =p_{1s}+p_{1+}\varepsilon _{p}e^{-i\delta
t}+p_{1-}\varepsilon _{p}^{\ast }e^{i\delta t},  \nonumber \\
&&\langle q_{2}\rangle =q_{2s}+q_{2+}\varepsilon _{p}e^{-i\delta
t}+q_{2-}\varepsilon _{p}^{\ast }e^{i\delta t},  \nonumber \\
&&\langle p_{2}\rangle =p_{2s}+p_{2+}\varepsilon _{p}e^{-i\delta
t}+p_{2-}\varepsilon _{p}^{\ast }e^{i\delta t},  \nonumber \\
&&\langle a\rangle =a_{s}+a_{+}\varepsilon _{s}e^{-i\delta
t}+a_{-}\varepsilon _{s}^{\ast }e^{i\delta t},
\end{eqnarray}
After substituting Eq. (4) into Eq. (3), and ignoring the
second-order terms, we obtain the steady-state mean values of the
system as
\begin{eqnarray}
&&p_{1s}=p_{2s}=0, \ \ \ \   q_{1s}=\frac{\hbar
g|c_{s}|^{2}}{m_{1}\omega
_{1}^{2}-\frac{\hbar^{2}\lambda ^{2}}{m_{2}\omega _{2}^{2}}},  \nonumber \\
&&q_{2s}=\frac{\hbar \lambda q_{1s}}{-m_{2}\omega _{2}^{2}}, \ \ \ \
a_{s}=\frac{\varepsilon _{l}}{i\Delta +\kappa },
\end{eqnarray}
with $\Delta =\Delta _{a}-gq_{1s}$ is the effective detuning of the
laser field from the cavity, and the solution of $a_+$
\begin{eqnarray}
a_+=\frac{1}{\kappa+i(\Delta-\delta)-\frac{i\hbar g^2
|a_s|^2}{A\times B}},
\end{eqnarray}
where
\begin{eqnarray}
&& A=m_1(\omega_1^2-\delta^2-i\delta\gamma_1)-\frac{\hbar^2\lambda^2}{m_2(\omega_2^2-\delta^2-i\delta\gamma_2)} , \nonumber \\
&& B=1+ \frac{i\hbar g^2 |a_s|^2}{A[\kappa-i(\Delta+\delta)]}.
\end{eqnarray}
Making use of the input-output relation of the cavity
\cite{QuantumOptics1994}, $\varepsilon _{out}(t)+\varepsilon
_{p}e^{-i\delta t}+\varepsilon _{l}=2\kappa \langle a\rangle$ and
$\varepsilon _{out}(t)=\varepsilon _{outs}+\varepsilon
_{out+}\varepsilon _{p}e^{-i\delta t}+\varepsilon _{out-}\varepsilon
_{p}^{\ast }e^{i\delta t}$, we can obtain $\varepsilon
_{out+}=2\kappa a_{+}-1$, which can be measured by homodyne
technique \cite{QuantumOptics1994}. Defining
$\varepsilon_R=\varepsilon _{out+}+1= 2\kappa a_{+}$, the reflective
output light $\varepsilon _R$ is of the same frequency $\omega _{s}$
as the signal field, which is a parameter in analogy to the
effective linear optical susceptibility. The real part of
$\varepsilon _R$ exhibits absorptive behavior, and its imaginary
part shows dispersive property. Although the cavity is empty, it can
be regarded as a material system composed of photons. As usual, we
determine the group velocity of light as
\cite{pra63-033804,pra46-29,oe17-19874},
\begin{eqnarray}
v_g=\frac{c}{n+\omega_s(\frac{dn}{d\omega_s})},
\end{eqnarray}
where the refractive index $n=1+2\pi \chi_{eff}$, and we can get
\begin{eqnarray}
 \frac{c}{v_g}=1+2\pi \texttt{Im}[\chi_{eff}(\omega_s)]_{\omega_s}+2\pi\omega_s
 \texttt{Im}[\frac{d\chi_{eff}(\omega_s)}{d\omega_s}]_{\omega_s},
\end{eqnarray}
here, $\chi_{eff}(\omega_s)$ is the effective susceptibility and is
in direct proportion to $\varepsilon_R$. Noticing that the signal
light has changed on the reflection and while
$\texttt{Im}[\chi_{eff}(\omega_s)]_{\omega_s}=0$, the group velocity
index should be written as
\begin{eqnarray}
 n_g=\frac{c}{v_g}=1+2\pi\omega_s\texttt{Im}[\frac{d\chi_{eff}(\omega_s)}{d\omega_s}]_{\omega_s}\propto
\texttt{Im}[\frac{\varepsilon_R(\omega_s)}{d\omega_s}]_{\omega_s}.
\end{eqnarray}
We can find from this expression  when the dispersion is steeply
positive or negative, the group velocity can be significantly
reduced or increased. In the following section we will present some
numerical results.
\section{Numerical results and discussion}
\begin{figure}
\centering
\includegraphics[width=8cm,height=5cm]{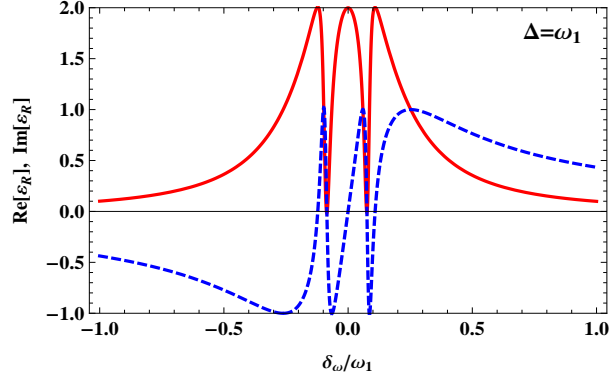}
\caption{\small {(Color online) The absorption $\texttt{Re}
[\varepsilon_R]$ (red solid line) and dispersion $\texttt{Im}
[\varepsilon_R]$ (blue dashed line) of the signal light as a
function of the $\delta_\omega/\omega_{1}=(\delta -\omega
_{1})/\omega_{1}$ while $\Delta=\omega_1$. The parameters used as
\cite{natcommun3-1196}, $\protect\lambda_l=$1064 nm, $d=$25 mm,
$\protect\omega_1=\protect \omega_2=2\protect\pi\times947\times10^3$
Hz, the quality factor
$Q_1=\frac{\omega_1}{\gamma_1}(Q_2=\frac{\omega_2}{\gamma_2})=6700$,
$m_1=m_2=145$ ng,
$\protect\kappa=2\protect\pi\times215\times10^3$Hz, $\wp_l=2$ mW,
and $\protect\lambda=8\times10^{35}$ Hz/m$^2$.}}
\end{figure}
For illustration of the numerical results, we choose the
realistically reasonable parameters to demonstrate the slow and fast
light effect based on the optomechanical system. We employ  the
parameters from the recent experiment \cite{natcommun3-1196}  in the
observation of the normal-mode splitting.

Figure 2 illustrates the behavior of the absorption
$\texttt{Re}[\varepsilon_R(\omega_s)]$  and dispersion
$\texttt{Im}[\varepsilon_R(\omega_s)]$ of the signal light  as a
function of $\delta_\omega/\omega_{1}=(\delta -\omega
_{1})/\omega_{1}$. We can find obviously that there are two steep
negative slopes  related to two minima of  zero-absorption in the
reflective light. The two minima of the zero-absorption in Fig.2 can
be evaluated by
$\frac{d\texttt{Re}[\varepsilon_R(\omega_s)]}{d\omega_s}\mid_{\omega_s=\omega_+,\omega_-}=0$,
where $\omega_+$ and $\omega_-$ are the points of the
zero-absorption minima.  This large dispersive characteristics can
lead to the possibility of implementation of fast light effect.
However, if we turn off the Coulomb coupling between the charged
NR$_1$ and NR$_2$, the two steep negative slopes disappear and only
one occurs. Which has been studied in Ref.\cite{pra87-013824}.
\begin{figure}[tbp]
\begin{minipage}[b]{0.5 \textwidth}
\includegraphics[width=1 \textwidth]{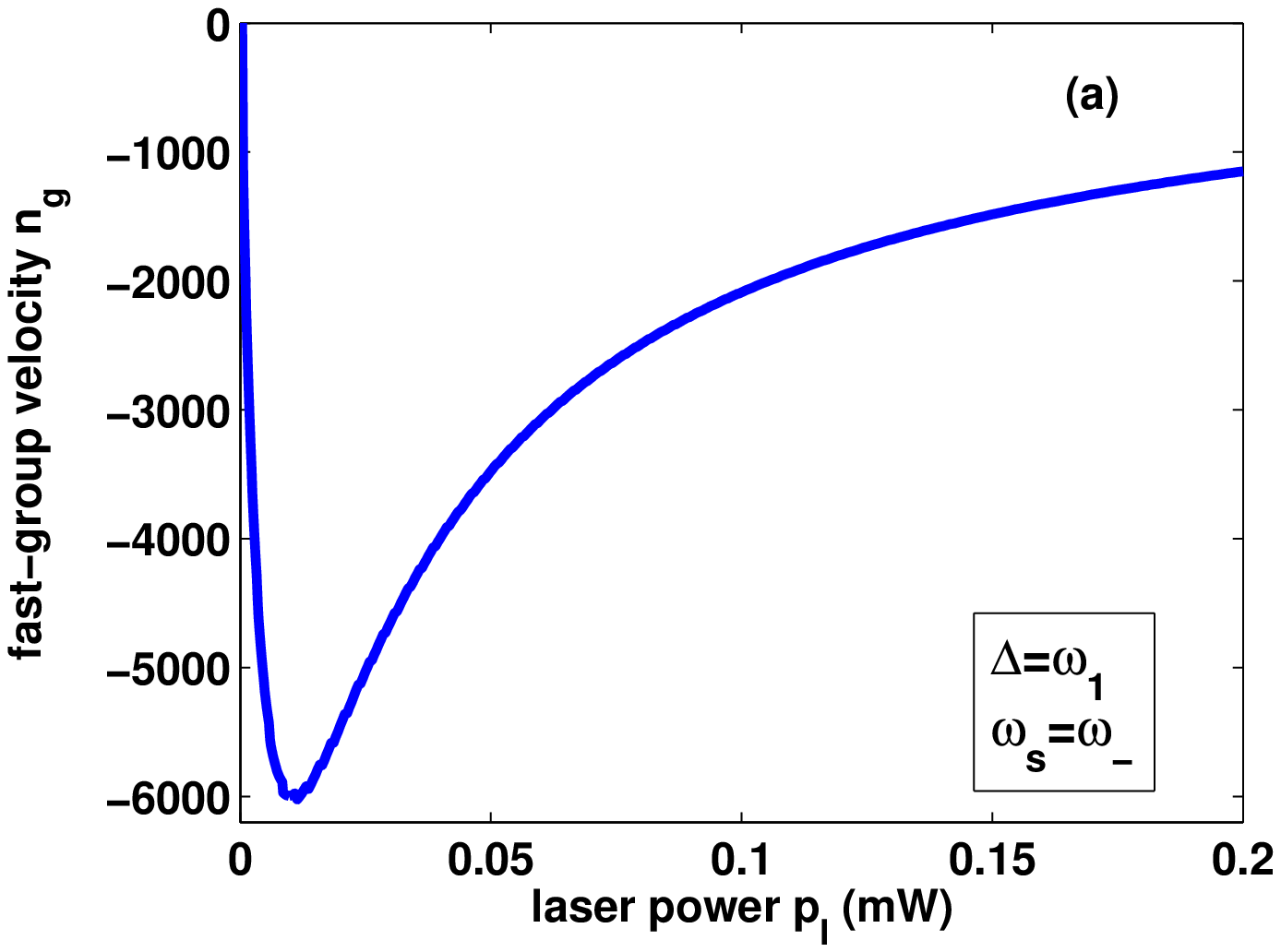}\includegraphics[width=1 \textwidth]{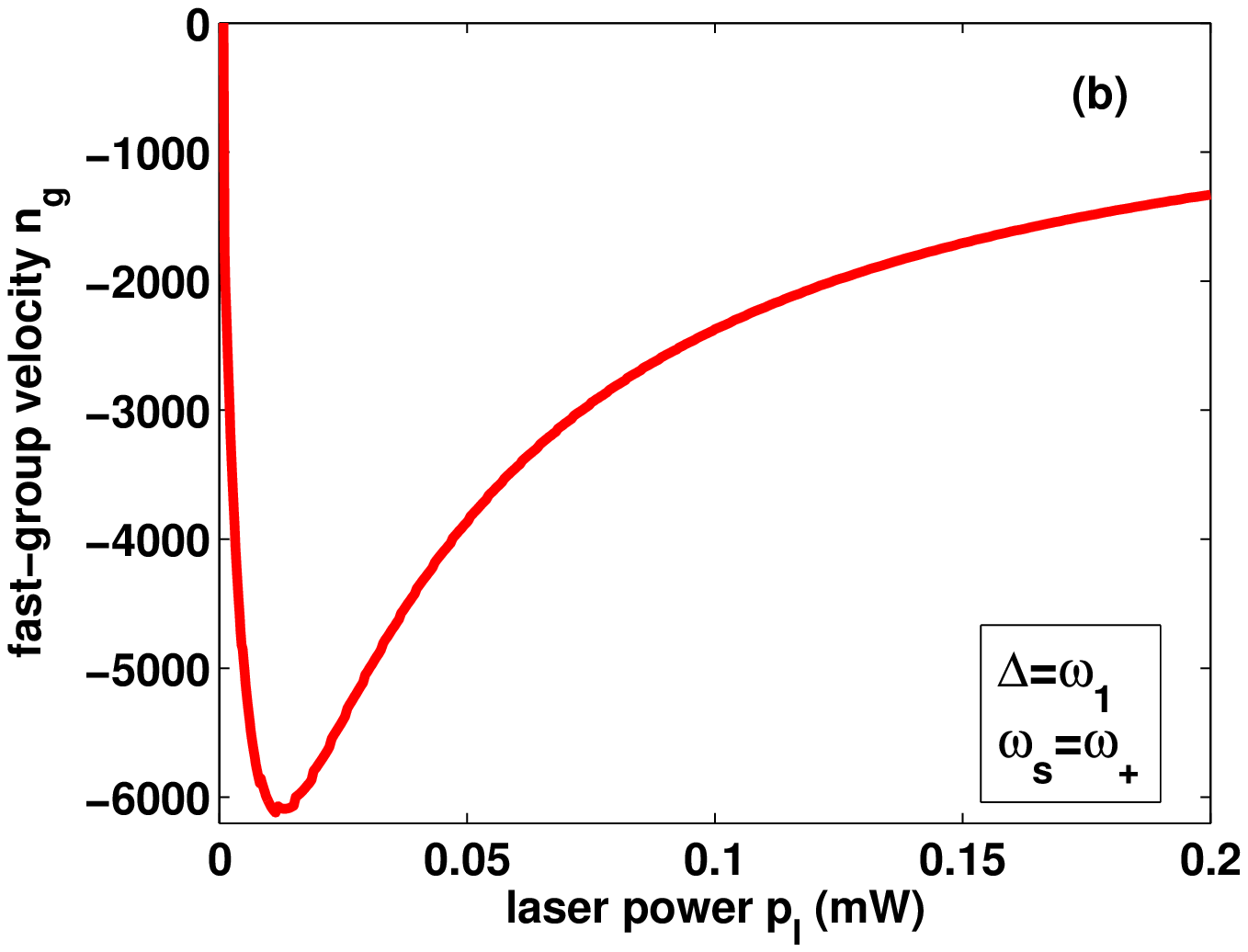}
\end{minipage}
\caption{\small {(Color online) The group fast  velocity index $n_g$
as a function of the laser power $\varepsilon_l$ while
$\Delta=\omega_1$, the parameters   used are the same as in Fig.
2.}}
\end{figure}

\begin{figure}
\centering
\includegraphics[width=8cm,height=5cm]{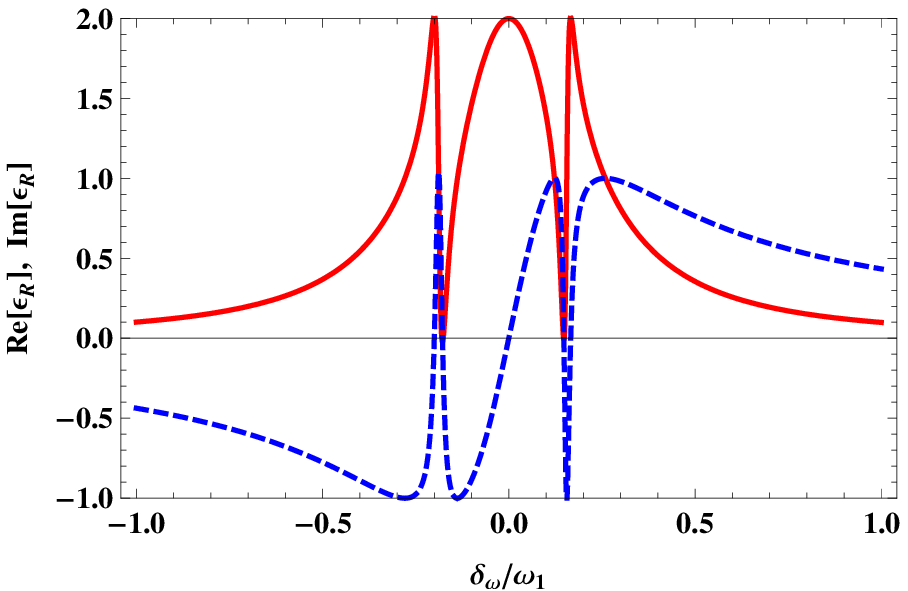}
\caption{\small {(Color online) The absorption
$\texttt{Re}[\varepsilon_R]$(red solid line) and dispersion
$\texttt{Im}[\varepsilon_R]$ (blue dashed line) of the signal light
as a function of the $\delta_\omega/\omega_{1}=(\delta -\omega
_{1})/\omega_{1}$ while Coulomb coupling strength
$\protect\lambda=2\times8\times10^{35}$ Hz/m$^2$. Other parameters
used are the same as in Fig. 2.}}
\end{figure}

Figure 3 shows the  two mode group   velocity $n_g$  as a function
of the laser power  and the parameters used are the same as in Fig.
2. It is clear that near $\varepsilon_l=0.01$mW,  the fast light
index can be obtained as 6000 times with two mode light frequencies
$\omega_+$ and $\omega_-$ respectively. That is, the output will be
6000 times faster than the input with two different frequencies.
Therefore, in our structure one can obtain the fast output light
without absorption by only adjusting the effective detuning of laser
field from the bare cavity equal to the frequency of the NR$_1$. The
physics of the effects can be explained by the radiation pressure
coupling an optical mode to a mechanical mode in an optomechanically
induced transparency (OMIT) \cite{science330-1520,nature472-69}. The
OMIT depends on quantum interference, which is sensitive to the
phase disturbance. The Coulomb coupling between the NR$_1$ and
NR$_2$ breaks down the symmetry of the OMIT interference, and thus
the single OMIT window is split into two OMIT
\cite{nat.phys.8-891,ma}.

Figure 4 describe the The absorption $\texttt{Re}[\varepsilon_R]$
and dispersion $\texttt{Im}[\varepsilon_R]$ of the signal light with
different Coulomb coupling strength. That is to say, the frequencies
of the signal light can be tuned by Coulomb coupling strength.
Similarly, the optomechanical system  also can implement the two
mode slow light effect without absorption when the effective
detuning of laser field from the bare cavity($\Delta=-\omega_1$). In
order to illustrate it more clearly, we plot Figs. 5 and 6 with the
same experimental data as in Fig. 2. In Fig. 5, we also describe the
theoretical variation of absorption $\texttt{Re}[\varepsilon_R]$ and
dispersion $\texttt{Im}[\varepsilon_R]$ of the signal light as a
function of $\delta_\omega/\omega_{1}=(\delta -\omega
_{1})/\omega_{1}$ when the detuning $\Delta=-\omega_1$. From Fig. 6
we can find that there are two the large dispersions  relates to two
very steep positive slopes. It means that there are two mode slow
light effect without absorption. Figure 6 shows the group velocity
index $n_g$ of  two mode slow light as a function of laser power.

\begin{figure}
\centering
\includegraphics[width=8cm,height=5cm]{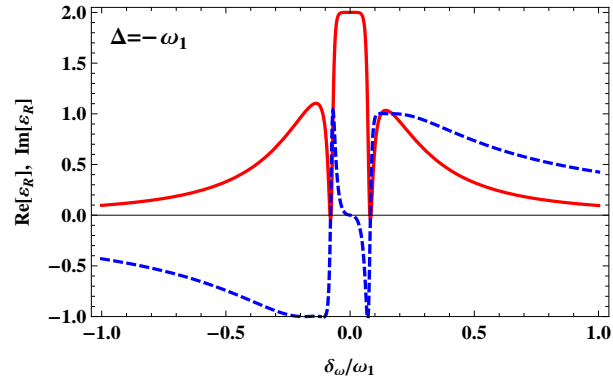}
\caption{\small {(Color online) The absorption
$\texttt{Re}[\varepsilon_R]$(red solid line) and dispersion
$\texttt{Im}[\varepsilon_R]$ (blue dashed line) of the signal light
as a function of the $\delta_\omega/\omega_{1}=(\delta -\omega
_{1})/\omega_{1}$ when $\Delta=-\omega_1$. Other parameters used are
the same as in Fig. 2.}}
\end{figure}

\begin{figure}[tbp]
\begin{minipage}[b]{0.5 \textwidth}
\includegraphics[width=1 \textwidth]{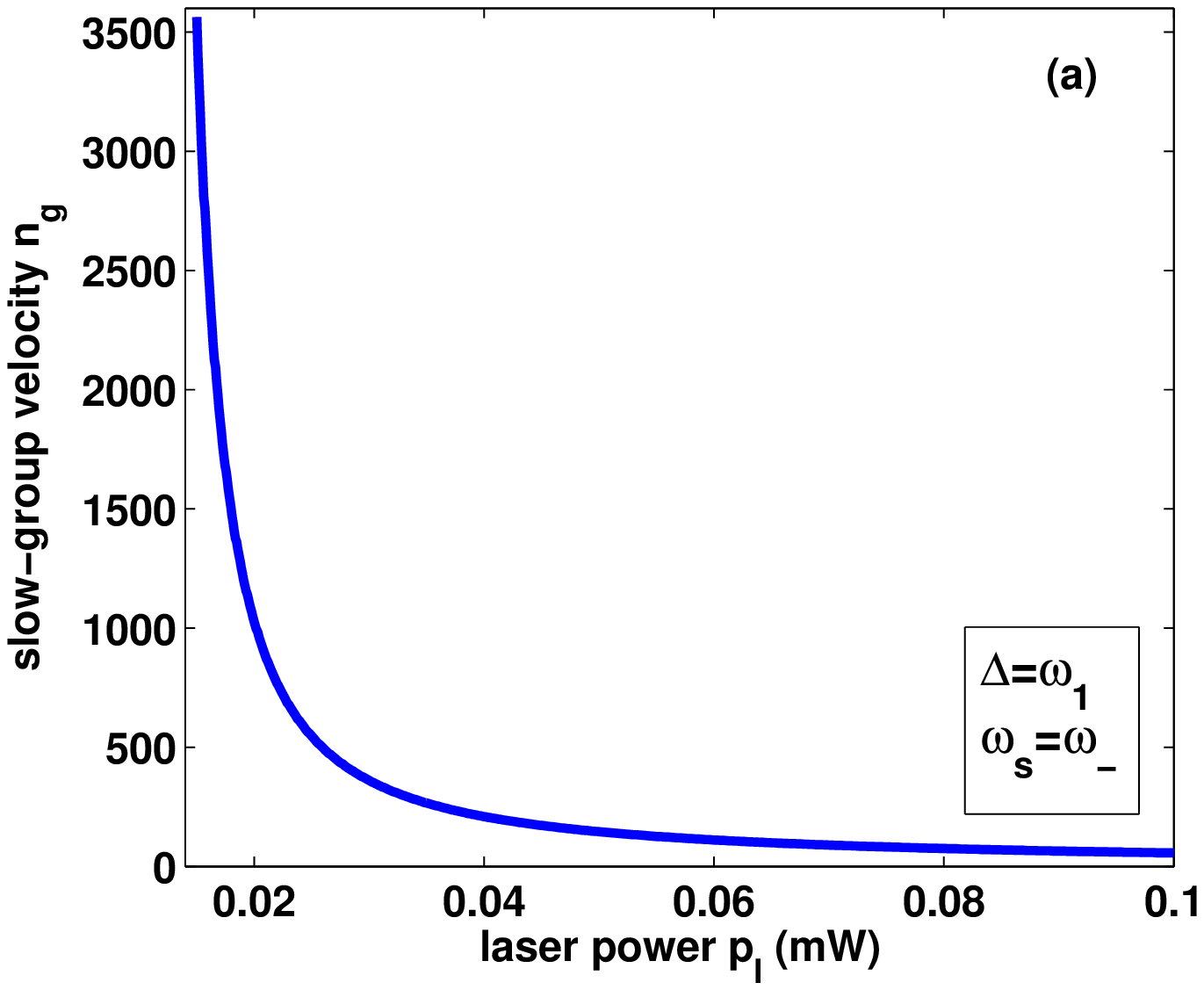}\includegraphics[width=1 \textwidth]{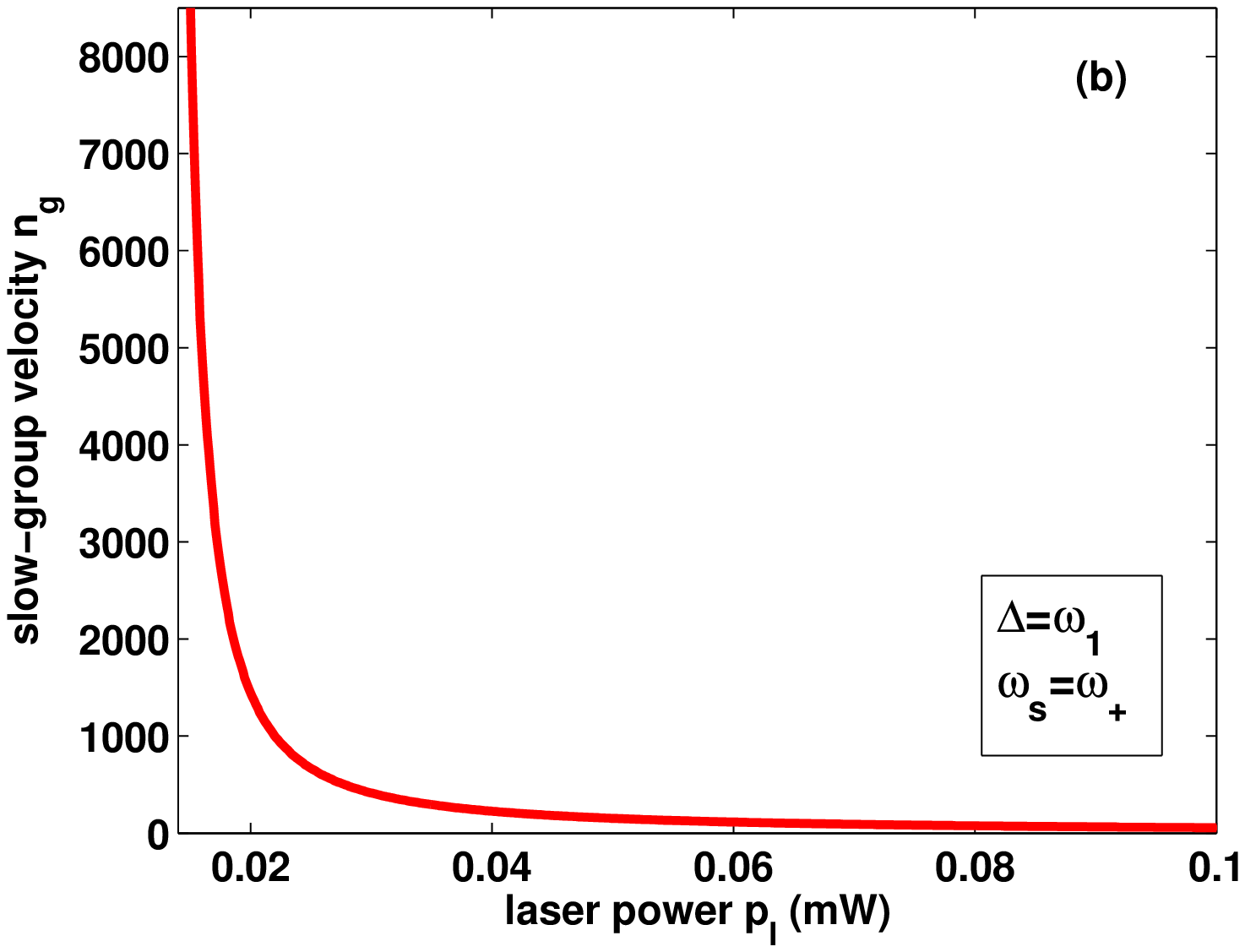}
\end{minipage}
\caption{\small {(Color online) The two mode slow  group  velocity
index $n_g$ as a function of the laser power $\varepsilon_l$ when
$\Delta=-\omega_1$. Other parameters   used are the same as in Fig.
2.}}
\end{figure}

According to the above discussions, it can be found clearly that the
optomenchanical system provides us an efficient and two mode  switch
between slow light and fast light by simply adjusting the effective
laser detuning in terms of the double-OMIT. In experiments, one can
fix the signal field with frequency $\omega_s=\omega_+, \omega_-$
and scan the effective  laser detuning from $\Delta=-\omega_1$ to
$\Delta=+\omega_1$, then one can efficiently switch the signal field
from slow to fast with two different frequencies.

\section{conclusions}
In conclusion, we have investigated  tunable fast- and slow- light
effects  in double mode with the optomechanical  system. It can
provide us an efficient and convenient way to switch between slow
and fast light with double-mode. The greatest advantage of our
system is that we can efficiently switch from fast to slow light by
only adjusting the laser-cavity  deturning. Moreover, the requencies
of signal light can be tuned by Coulomb coupling strength. Our
scheme may have potential applications in various applications such
as optical communication, nonlinear optics.

Finally, we hope that the results of this paper can be tested by
experiments in the near future. Recently, S. Weis et
al.\cite{science330-1520} and A. H. Safavi-Naeini et
al.\cite{nature472-69}  have reported experimental results on
signal-mode slow light and OMIT in a optomechanical system, maybe
one can use a similar experimental setup to test our predicted
effects.
\section*{ACKNOWLEDGMENTS}
This work was supported by the Natural Science Funding for Colleges
and Universities in Jiangsu Province (Grant No. 12KJD140002), and
Program for Excellent Talents of Huaiyin Normal University(No.
11HSQNZ07).

\end{document}